\documentclass[aps,twocolumn,showpacs,amsmath,amssymb]{revtex4}
\usepackage{dcolumn}
\usepackage{bm}
\usepackage{color}
\usepackage{latexsym}
\usepackage{amssymb}
\usepackage{graphicx}
\usepackage{ulem}
\begin{document}
\title{Resonance terahertz detection in ungated two-dimensional electron gas}

\author{M. V. Cheremisin }

\affiliation{A.F.Ioffe Physical-Technical Institute, 194021
St.Petersburg, Russia}

\begin{abstract}
The response of an ungated two-dimensional electron gas(2DEG) to an external electromagnetic excitation is analyzed. The possibility of creating a single-mode resonance detector operating in the terahertz frequency range is demonstrated.
\end{abstract}


\maketitle

\section{\label{sec:Introduction}Introduction}
The plasma wave generation mechanism [1] in an ultrashort-channel Field Effect Transistor
(FET) has attracted a considerable attention. The high-density 2DEG can be used for detection [2],[3] of high-frequency radiation as well. The generation [4],[5] and both the non-resonant [6] and resonant [7]-[9] detection in the terahertz (THz) frequency range(see also [10]) have been demonstrated experimentally.

Recently, it has been shown that the current instability and the resulting plasma wave generation similar
to that in FET can be obtained in ungated 2D electron layers [11]. Moreover, the possibility of a single-mode THz generator based on an ungated 2D electron layer was discussed in Ref.[12]. In the present paper, we calculate the response of a detector based on ungated 2D electron layer for an arbitrary current on the assumption of a dissipationless carrier transport. The role of the finite carrier scattering
strength is discussed. We demonstrate the feasibility of a single-mode detector operating in the
terahertz frequency range.

\section{\label{sec:Analytical approach}Analytical approach}

We use the hydrodynamic model of a weakly damped, incompressible charged-electron fluid
placed in a rigid, neutralizing positive background. The mean free path associated with electron-electron
collisions is assumed to be less than that related to scattering by phonons and
impurities and than the device length. The behavior of the 2D
fluid is described [11] by the Euler and the
continuity equations:

\begin{eqnarray}
\frac{\partial v}{\partial t}+v\frac{\partial v}{\partial
x}=-\frac{e}{m}\frac{\partial \phi}{\partial x}-\frac{v}{\tau},
\label{Euler} \\
\frac{\partial n}{\partial t}+\frac{\partial(vn)}{\partial x}=0.
\label{continuity}
\end{eqnarray}
Here, $n$ is the 2D electron density; $v$, electron
flux velocity; $m$, effective mass; $-e$, electronic
charge; and $\phi$, potential, so that the electric field
is $E=-\partial \phi/\partial x$. Let us analyze the detector mode for the equivalent circuit shown
in Fig.1. The boundary conditions at the source($x=0$) and drain($x=l$) are
\begin{figure}[tbp]\vspace*{0.5cm}
\includegraphics[scale=0.7]{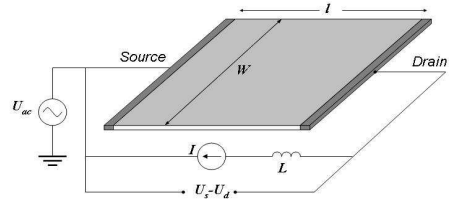}
\caption[]{\label{Fig1} Ungated 2D electron gas detector setup configuration. The response $\Delta U=U_{s}-U_{d}$ is caused by the incident radiation $U_{a}$.} \vspace*{-0.5cm}
\end{figure}

\begin{eqnarray}
n(0,t)=n_{0}+n_{a}cos(\omega t),\nonumber \\
n(l,t)v(l,t)=j/We, \label{boundary conditions}
\end{eqnarray}
where $j$ is the current fixed at the drain, and $W$ is the 2D layer width. We further suppose that the incident electromagnetic radiation with a wavelength substantially longer than the device dimensions is collected by an antenna coupled with 2DEG device. The boundary condition at the source can be realized by implementing the antenna output between the source contact and the ground. Therefore, the background 2DEG density at the source $n_{0}$ is kept constant. Then, the slot antenna provides an ac voltage $U_{ac}=U_{a}\cos(\omega t)$, and, hence the subsequent small variation of the 2DES density at the source $n_{ac}=n_{a}\cos(\omega t)$. Finally, connecting the drain to the power supply via an inductance provides that the current at the drain is maintained constant.

In the absence of radiation, the steady state corresponds to a current-carrying electron fluid with a constant density $n=n_{0}$( because of the local quasi-neutrality of the 2D fluid ) and constant flux velocity $v_{0}=\mu E=j/eWn_{0}$, where $\mu=e\tau/m$ is the carrier mobility. The steady-state potential $\phi_{0}$ is linear downstream the electron flow. In the absence of dissipation $\phi_{0}=0$.

We now seek the solutions to Eqs.(1) and
(2) in the following form

\begin{eqnarray}
n=\overline {n} + n_{1}(x,t), \nonumber \\
v=\overline {v} +n_{1}(x,t), \nonumber \\
\phi=\overline {\phi} + \phi_{1}(x,t), \label{perturbations}
\end{eqnarray}
where the time-dependent variations $v_{1}(x,t), n_{1}(x,t)$ are small, compared to the electron fluid velocity $\overline {v}$ and 2DEG density $\overline {n}$ time-averaged over the period $2\pi/\omega$ . For a small input signal amplitude, $n_{a}$, the perturbations $v_{1}, n_{1}, \phi_{1} $ are proportional to $n_{a}$, while $\overline {n} -n_{0}, \overline {v}-v_{0}$ are proportional to $n_{a}^{2}$. To the first order in $n_{a}$, we obtain

\begin{eqnarray}
\frac{\partial v_{1}}{\partial t}+v_{0}\frac{\partial v_{1}}{\partial x}+\frac{e}{m}\frac{\partial \phi_{1}}{\partial x}+\frac{v_{1}}{\tau}=0,
\nonumber \\
\frac{\partial n_{1}}{\partial t}+n_{0}\frac{\partial v_{1}}{\partial x}+v_{0}\frac{\partial n_{1}}{\partial x}=0.
\label{first order eqs}
\end{eqnarray}

Retaining time-independent terms of the second order in $n_{a}$ in Eq.(1) and Eq.(2), we find

\begin{eqnarray}
\frac{\partial}{\partial
x}\left[ \frac{1}{2}\overline {v_{1}^2} + \frac{e \overline {\phi}}{m} \right]+\frac{\overline {v}}{\tau}=0,
\nonumber \\
\frac{\partial}{\partial
x}\left[ n_{0} \overline {v} + \overline{ n_{1} v_{1}} \right]=0.
\label{second order eqs}
\end{eqnarray}

Integration of Eq.(6)
with respect to $x$ yields the dc detector response $\Delta \phi=\overline {\phi(l)}- \overline {\phi(0)}$:

\begin{equation}
\Delta \phi=\frac{m}{e} \left ( \frac{\overline {v_{1}(0)^2}}{2}-\frac{\overline {v_{1}(l)^2}}{2} + \frac{1}{\tau} \int _{0}^{l} \frac{\overline{ n_{1} v_{1}}}{n_{0}} dx  \right )+ \frac{m v_{0}l}{e \tau },
\label{detector response}
\end{equation}
where the last term corresponds to the conventional ohmic voltage drop. Looking for an explicit expression for the dc detector response given by Eq.(7), we now solve Eq.(5) for small perturbations $v_{1}, n_{1},\phi_{1} \sim \exp(-i\omega t+ikx)$ and then obtain

\begin{eqnarray}
(\omega-kv_{0})v_{1}=-\frac{e}{m}k\phi_{1}-\frac{i}{\tau}v_{1}, \nonumber \\
(\omega-kv_{0})n_{1}=kn_{0}v_{1}. \label{linearizing}
\end{eqnarray}
Neglecting the finite-size effects [13], we explore the
relation $\phi_{1}=-\frac{2\pi e n_{1}}{\left |k \right |
\epsilon}$ known for infinite ungated 2D electron layer, where $\epsilon$ is the background dielectric constant. Therefore, the dispersion equation yields:

\begin{equation}
(\omega-kv_{0})\left (\omega-kv_{0}+\frac{i}{\tau}\right )=2a\left |k \right |,
\label{dispersion}
\end{equation}
where $a=\frac{\pi n_{0}e^{2}}{\epsilon m}$. Introducing the
dimensionless frequency
$\Omega=\omega \sqrt{\frac{l}{a}} $, scattering parameter $\gamma=\frac{1}{\tau}\sqrt {\frac{l}{a}}$, and flux velocity $V=v_{0}/\sqrt{la}$,
we derive the dispersion relation for plasma waves propagating
downstream $k_{+}$ and upstream $k_{-}$ the current flow as
follows:

\begin{equation}
k_{\pm}=\pm \frac{1}{l}\frac{1 \pm (\Omega+i\frac {\gamma}{2}) V-\sqrt{1 \pm
2(\Omega+i\frac {\gamma}{2}) V-\frac{V^{2}\gamma^{2}}{4}}}{V^{2}}.
\label{wave vectors}
\end{equation}
For dissipationless carrier transport $\gamma=0$ and small current flux, Eq.(10) reproduces the dispersion relation
$k_{\pm}=\pm \frac{\omega^{2}}{2a}\left (1 \mp \frac{\omega v_{0}}{a}\right )$
found previously in Ref.[11]. Then, in the absence of current, Eq.(10) corresponds to the dispersion relationship $k_{\pm}=\pm \frac{\omega^{2}}{2a}(1 + \frac{i}{\omega \tau})$ for a free plasmon in an infinite 2D media.

Seeking the solution to
Eq.(8) in the form $n_{1}=C_{+}
 \exp(i k_{+}x)+C_{-} \exp(i k_{-}x)$ and, then using the boundary
conditions described by Eq.(3), we immediately obtain

\begin{eqnarray}
n_{1} = Re \left [ (C_{+}e^{ik_{+}x}+C_{-}e^{ik_{-}x})e^{-i \omega t} \right ],\nonumber \\
v_{1} = Re \left [ \left ( C_{+} \frac{w-k_{+}v_{0}}{n_{0}k_{+}} e^{ik_{+}x} + C_{-} \frac{w-k_{-}v_{0}}{n_{0}k_{-}} e^{ik_{-}x} \right ) e^{-i \omega t} \right ], \nonumber \\
C_{\pm}=\frac{n_{a}}{1-\frac{k_{\pm}}{k_{\mp}}e^{i(k_{\pm}-k_{\mp})l}}. \nonumber \\
\label{linear solution}
\end{eqnarray}
We can now find the detector response specified by Eq.(7). Nevertheless, we attempt to  qualitatively describe the detector response by using the results of Refs.[11],[12] for the plasma wave instability in an ungated 2DEG fluid. It is well known that the response of a resonance-like system to an external harmonic excitation is strongly correlated with eigenvalues of the energy spectrum. As it was demonstrated in Ref.[11], the steady-state electron flow is unstable without external excitation, i.e., at $n_{a}=0$. The ungated 2DEG layer can be viewed as a resonator cavity with the complex frequency spectrum  $\Omega_{N}=\Omega'_{N}+i\Omega''_{N}$, where $N=1,2..$ is the mode index. The $N$-th mode is unstable when the instability increment is positive, i.e., at $\Omega''_{N}>0$. For undamped plasma waves $\Omega'_{N} \gg \Omega''_{N}$ the complex frequency was calculated in Ref.[12]

\begin{eqnarray}
\Omega'_{N}=\frac{1}{2V}\left[1-\left(\frac{(2-\pi V^{2}(2N-1))^{2}-2}{2}\right)^{2} \right]^{1/2}, \label{frequency} \\
\Omega''_{N}=V\frac{\sqrt{1-4(\Omega'_{n}V)^{2}}}{\sqrt{1-2\Omega'_{n}V}-\sqrt{1-2\Omega'_{n}V}}\ln
\left|R_{s}R_{d} \right |-\frac{\gamma}{2}, \label{increment} \\
R_{s}=-1,
R_{d}=\frac{1-\Omega'_{n}V-\sqrt{1-2\Omega'_{n}V}}{1+\Omega'_{n}V-\sqrt{1-2\Omega'_{n}V}}. \nonumber
\end{eqnarray}
For a small current, Eqs.(12),(13) describe [11] the discreet
frequency spectrum $\omega'_{N}=\sqrt{\pi a(2N-1)/l}$ and
mode-independent instability increment $\omega''=v_{0}/l-1/2\tau$.

Under current-carrying conditions, the complex frequency exhibits unexpected features discussed in detail in Ref.[12]. In Fig.2, we plot both the frequency and increment for three lowest modes as a function of the flux velocity for dissipationless carriers $\gamma=0$. The $N$-th mode frequency exhibits a well pronounced red-shift upon an increase in the current. Then, each mode is characterized by the respective increment, which is positive within a certain range of flux velocities $0<V<V^{cut}_{N}$, where $V^{cut}_{N}=\sqrt{\frac{2-\sqrt{2}}{\pi (2N-1)}}$ is the cutoff velocity of the $N$-th mode. Actually, the current-driven cutoff of the instability is caused by vanishing of the upstream plasma wave group velocity $d\omega/dk_{-}$ when the relation $\Omega'=1/2V$ is valid(see Eq.(10) at $\gamma=0$). We state that only the first mode remains unstable at high currents $V>V^{cut}_{2}=0.25$.
\begin{figure}[tbp]\vspace*{0.5cm}
\includegraphics[scale=0.7]{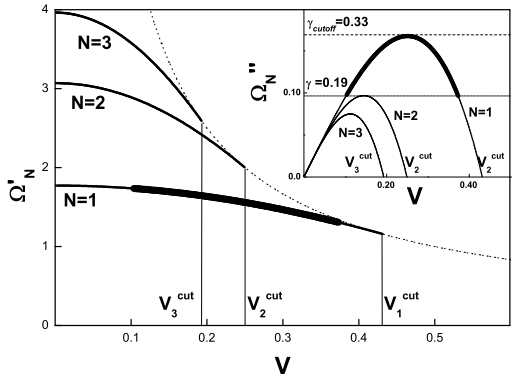}
\caption[]{\label{Fig2} Dimensionless frequency $\Omega'_{N}$(main panel) and increment $\Omega''_{N}$ (inset) vs. the
dimensionless velocity $V$ for dissipationless carriers for
$\gamma=0$ and $N=1,2,3$ modes. The dotted line in the main panel represents the instability cutoff relation $\Omega'=1/2V$. The bold line represents the frequency and the increment (inset) of the single-mode excitation in the presence of a finite dissipation $\gamma=0.19$. }\vspace*{-0.5cm}
\end{figure}
For the actual case of finite scattering ($\gamma \neq 0$), the instability increment is independent of the mode index(see Eq.(13)), and, therefore, all the modes are damped identically. It is easy to check whether $N$-th mode is still present in the plasma wave spectrum by plotting the damping strength $\gamma/2$ in the instability increment plot( see Fig.2, inset). Above a certain critical value of dissipation $\gamma_{cr} >0.33$, the instability disappears. Then, for $0.19 <\gamma < \gamma_{cr}$, only the proper mode remains unstable. Note that, at small dissipation $\gamma<0.19$, an increase in the current must result in a step-by-step suppression of, first, higher- and, then, lower-index modes.

Now we can predict some essential features of the detector response. At $\omega \tau=\Omega/\gamma \gg 1$, the response is expected to be resonant at the proper- and higher-order mode frequencies given by Eq.(12). For intermediate values of scattering $\gamma < \gamma_{cr}$, an increase in the current must lead to suppression of, first, higher- and, then, lower-index resonator modes. For certain values of $\gamma, V$, the detector response must  exhibit a single-mode behavior. In the opposite case of a strong dissipation $\Omega/\gamma \leq 1$, the response becomes non-resonant.  In order to confirm our predictions, we analyze in more detail the detector response in two limiting cases: a) $V \neq 0, \gamma=0$ and b) $V=0, \gamma \neq 0$.

\subsection{\label{subsec: Selective detector}Selective detector}
We now find the detector response under the current-carrying condition $V \neq 0$ and zero dissipation $\tau=\infty$. Substituting Eq.(11) into Eq.(7), we obtain the detector response
$\Delta \phi=\frac{m}{2e} \left ( \overline {v_{1}(0)^2}-\overline {v_{1}(l)^2} \right )$ as follows

\begin{eqnarray}
\Delta \phi = \frac{mal}{e}\left ( \frac{n_{a}}{n_{0}} \right )^{2} F(\Omega), \nonumber \\
F(\Omega)=\frac{(\Omega-\kappa_{+}V)(\Omega-\kappa_{-}V)\sin^{2}({\frac{\kappa_{+}-\kappa_{-}}{2}})}{(\kappa_{+}-\kappa_{-})^{2}+4\kappa_{+}\kappa_{-}\sin^{2}({\frac{\kappa_{+}-\kappa_{-}}{2}})},
\label{detector response(current)}
\end{eqnarray}
where we introduced dimensionless wave vectors $\kappa_{\pm}=k_{\pm}l$ and $F(\Omega)$ is the response function. At small frequencies $\Omega \ll 1$ the detector response is described by the asymptote $F(\Omega) \sim \Omega^{2}/4$, and thus is independent of the current. Then, at a zero drain current, Eq.(14) describes the detector response $F(\Omega)= \frac{\tan^{2}(\Omega^{2}/2)}{\Omega^{2}}$ which exhibits sharp resonances(see Fig.3) at the fundamental frequency and its odd harmonics $\Omega_{N}=\sqrt{\pi(2N-1)}, N=1,2..$. As expected, in the opposite case of a nonzero current, the response function specified by Eq.(14) demonstrates the noticeable red-shift of the resonator peaks(see Fig.3). At a fixed carrier flux velocity $V$, the detector response exhibits a cutoff at a certain frequency $\Omega'=\frac{1}{2V}$. When $V>V^{cut}_{2}=0.25$, only the first mode of the resonator can be excited, and, thus, the device operates as a single-mode detector(see heavy curve in Fig.3).
\begin{figure}[tbp]\vspace*{0.5cm}
\includegraphics[scale = 0.7]{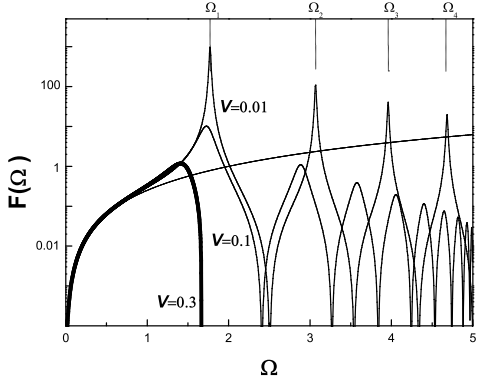}
\caption[]{\label{Fig3} Detector response function specified by Eq.(14) vs. the dimensionless frequency at $V=0.01,0.1$ and $0.3$(bold line) for dissipationless carriers $\gamma=0$. Thin line shows the asymptote at $\Omega \ll 1$} \vspace*{-0.5cm}\vspace*{-0.5cm}
\end{figure}
\subsection{\label{subsec:Dissipation} Dissipation induced damping of the detector response}
We are now interested in the case of a detector operating under zero-current conditions $V=0$ and finite dissipation $\gamma \neq 0$. The dispersion relationship is simplified as follows $\kappa_{\pm}=\pm \frac{\Omega}{2} \left( \Omega+i\gamma \right )$. The detector response given by Eq.(7) yields
\begin{eqnarray}
F(\Omega)=\frac{1}{\gamma^{2}+\Omega^{2}} \frac{2\sinh^{2}\left (\frac {\Omega \gamma}{2}\right )+(1-\frac{\gamma^{2}}{\Omega^{2}})\sin^{2}\left (\frac{\Omega^{2}}{2}\right ) }{\sinh^{2}\left (\frac {\Omega \gamma}{2}\right )+\cos^{2}\left (\frac{\Omega^{2}}{2}\right )}.
\label{detector response(dissipation)}
\end{eqnarray}
Without dissipation, i.e. at $\gamma=0$ Eq.(15) we reproduce the above-mentioned result $F(\Omega)= \frac{\tan^{2}(\Omega^{2}/2)}{\Omega^{2}}$ which yields the zero-frequency asymptote as $\Omega^{2}/4$. An increase in the dissipation results in a smoothing of the resonant peaks (see Fig.4) which then shrink when the detector quality factor becomes on the order of unity, i.e., $\omega \tau=\Omega/\gamma \sim 1$. In the opposite case of a strong dissipation and(or) long sample $k''l=\Omega \gamma/2 \gg 1$ the oscillations excited at the source by the incident radiation do not reach the drain because of the damping. The boundary condition at the drain is irrelevant in this case and the response is independent of the sample length since $F(\Omega)=\frac{2}{\gamma^{2}+\Omega^{2}} \sim 1/l$. At substantially higher frequencies $\Omega \gg \gamma$, the response becomes independent of the scattering strength $F(\Omega)=2/\Omega^{2}$.

\begin{figure}[tbp]\vspace*{0.5cm}
\includegraphics[scale = 0.7]{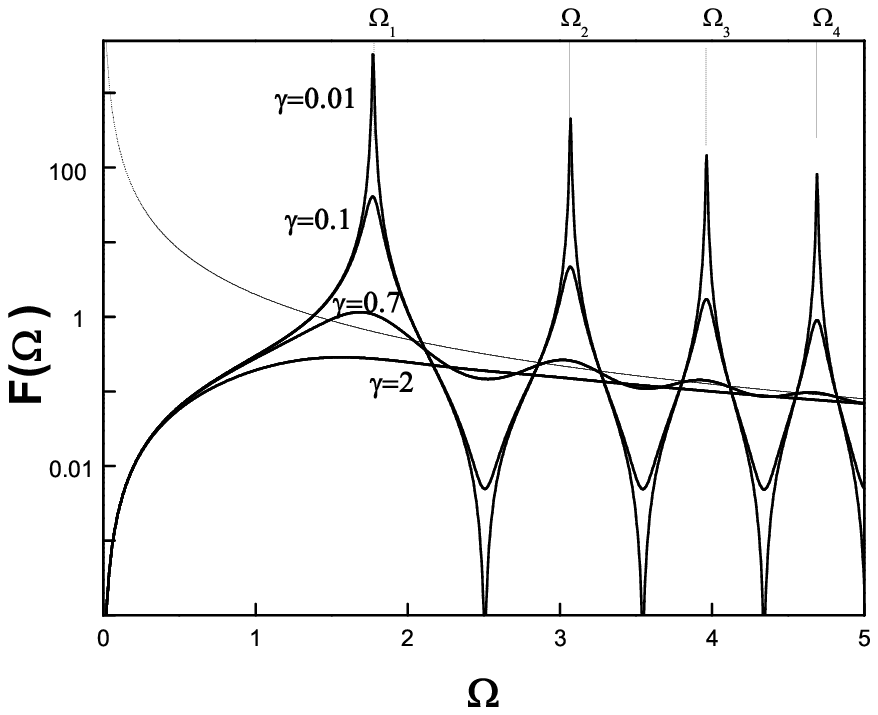}
\caption[]{\label{Fig4} Detector response function specified by Eq.(15) vs. the dimensionless frequency at $\gamma=0.01,0.1,0.7,2 $ for $V=0$. Dashed line represents the asymptote at $\Omega \gg 1$} \vspace*{-0.5cm}\vspace*{-0.5cm}
\end{figure}

\section{\label{subsec:Discussion} Discussion}
Let us estimate the values of the parameters determined in the present paper. For $l=1\mu$m channel and the typical 2DEG  density $n_{0}=10^{12}$sm$^{-2}$, effective mass $m=0.042m_{0}$, and dielectric constant $\epsilon=14$ of InGaAs lattice matched with InP, we obtain the proper mode frequency $\omega_{1}=6.5$THz. Then, assuming the 2DEG mobility $\mu=10^{5}$V/sm s, we find the relaxation time $\tau=3.8 \times 10^{-12}$s and the scattering parameter $\gamma=0.07$. Therefore, the proper mode is undamped since the quality factor $\omega\tau=25 \gg 1$. The device operates as a single-mode detector when the carrier flux velocity exceeds the value $v_{0}=V_{2}^{cut} \sqrt{al}=9.3 \times 10^{5}$m/s.

We now estimate the detector responsivity defined in Ref.[14].
\begin{equation}
R=\frac{\Delta \phi}{SI},
\label{responsivity1}
\end{equation}
where $S=\lambda^{2}G/4\pi$ is the antenna aperture; $G \simeq 1.5$, dipole antenna gain factor; $\lambda$,  electromagnetic radiation wavelength; $I=\frac{cE^{2}}{8\pi}$, electromagnetic field intensity; $E$, ac field amplitude of the electromagnetic wave in a vacuum. Substituting Eq.(14) into Eq.(16) and estimating the ac voltage as $U_{a}=E\lambda/4$, we finally obtain
\begin{equation}
R=R_{0}F(\Omega), R_{0}=\frac{\pi^{3}\epsilon}{8Gceln_{0}},
\label{responsivity2}
\end{equation}
where we use the relationship $U_{a}=-\frac{2\pi e n_{a}}{\left |k \right | \epsilon}$ and, moreover, assume $k=\pi/2l$ for actual first mode detection case. For channel length $l=1\mu$m, dielectric constant $\epsilon=14$, typical 2DEG  density $n_{0}=10^{12}$sm$^{-2}$, we obtain $R_{0}= 76$[V/W]. Since the peak magnitude of the response function $\sim 10^{3}$, the detector responsivity is on the order of $\sim 10^{5}$[V/W] which far exceeds the typical values for Schottky diode detectors(on the order of $\sim 10^{3}$[V/W]).

\section{\label{subsec:Conclusions} Conclusions}
In conclusion, we demonstrate that a short-channel ungated 2D electron gas exhibits a resonance response to an incident electromagnetic radiation in the terahertz frequency range. The responsivity of the 2DEG-based detector exceeds that of Schottky diodes. We demonstrate the possibility of a single-mode resonance selective detector operating at terahertz frequencies.

\end{document}